\documentclass[12pt,preprint]{article}

\usepackage{amsmath}

\usepackage{graphics}

\usepackage{epsfig}

\renewcommand{\vec}[1]{{\bf #1}}

\usepackage{bbm}

\newcommand{\eqb}{\begin{equation}}
\newcommand{\eqe}{\end{equation}}
\newcommand{\dmb}{\begin{displaymath}}
\newcommand{\dme}{\end{displaymath}}

\newcommand{\eab}{\begin{eqnarray}}
\newcommand{\eae}{\end{eqnarray}}

\newcommand{\be}{\begin{equation}}
\newcommand{\ee}{\end{equation}}

\begin{document}

\begin{titlepage}

\begin{flushright} 
KA-TP-12-2008
\end{flushright}

\vspace{0.6cm}

\begin{center}

\Large{Thermal photon dispersion law and modified black-body spectra}

\vspace{1.5cm}

\large{Josef Ludescher$\mbox{}^\dagger$ and Ralf Hofmann$\mbox{}^*$}

\end{center}

\vspace{1.5cm}

\begin{center}

{\em $\mbox{}^\dagger$Institut f\"ur Theoretische Physik\\ 
Universit\"at Heidelberg\\ 
Philosophenweg 16\\ 
69120 Heidelberg, Germany\vspace{0.5cm}\\ 
$\mbox{}^*$Institut f\"ur Theoretische Physik\\ Universit\"at Karlsruhe
(TH) \\ 
Kaiserstr. 12\\ 
D-76131 Karlsruhe, Germany}

\end{center}

\vspace{1.5cm}

\begin{abstract}

Based on the postulate that photon propagation is governed by an SU(2)
gauge principle we numerically compute the one-loop 
dispersion for thermalized photon propagation on the radiatively
induced mass shell. Formerly, the dispersion was 
addressed by assuming $p^2=0$. While this approximation turns out to be excellent 
for temperatures $\le 2\,T_{\tiny\mbox{CMB}}$ the exact 
result exhibits a much faster (power-like) shrinking 
of the gap in the black-body spectral intensity with rising 
temperature. Our previous statements on anomalous large-angle 
CMB temperature-temperature correlations, 
obtained in the approximation $p^2=0$, remain valid. 

\end{abstract}

\end{titlepage}

\section{Introduction}

The possibility that photon propagation is governed by an SU(2) rather
than a U(1) gauge principle\footnote{For a feasibility discussion of this
postulate see \cite{GH2005}.} would have exciting astrophysical, cosmological,
and conceptual consequences. To make them quantitative a precise 
understanding of SU(2) gauge dynamics is required. For the physics of
photon propagation an analytical grasp of the deconfining phase is
sufficient. Thermodynamically, this is feasible 
\cite{Hofmann2005,HH2004,Hofmann2007,GH2006,Hofmann2006,KH2007}.  

In \cite{SHG2006} we have computed the one-loop polarization tensor
$\Pi$ of the
massless mode in the deconfining phase of thermalized 
SU(2) Yang-Mills theory. By numerically finding estimates on the moduli
of the one-particle irreducible contributions to the pressure at three-loop
level \cite{KH2007} we conclude that the associated two-loop contributions 
to $\Pi$, obtained by cutting a massless line in the corresponding diagram
for the pressure, are extremely suppressed with respect to 
the one-loop result (by a factor of the order $10^{-3}$). 
This entails that a modification of the massless mode's 
dispersion law by radiative effects is in the effective 
theory for all practical purposes exhaustively described at the
(resummed) one-loop level.  

To be able 
to explicitely solve the kinematic constraints for the integration
variables in the one-loop expression for $\Pi$, which are imposed by 
a maximal, temperature-dependent resolution scale $|\phi|$ owing to a 
thermal ground state\footnote{The notion of a thermal
  ground state emerges upon the execution of a sufficiently local, 
selfconsistent spatial coarse-graining over interacting calorons and anticalorons of
topological charge modulus unity \cite{HH2004,Hofmann2007}.}, we have so far 
resorted to the approximation that the square of the four momentum $p$ of 
the massless mode be zero: $p^2=0$. In this approximation, there is no
imaginary contribution to the screening function $G$. For temperatures not far
above the critical temperature $T_c$ for the deconfining-preconfining
phase transition, $G$ exhibits a sizable small-momentum regime where
$G>\vec{p}^2$. This regime is associated with a strong screening of the 
massless mode. On the other hand, there is a momentum regime where $G$
is sizably negative (antiscreening) thus describing the average 
loss of the massless mode's energy to the ground-state
dynamics. Microscopically, this effect is related to the generation of
stable and nonrelativistic monopole-antimonopole pairs out of
small-holonomy (anti)calorons. The latter's holonomy is shifted to large values
by inelastic scattering of the massless mode off the originally short-lived 
monopole-antimonopole systems (caloron or anticaloron dissociation). The
rare generation of stable magnetic matter by scattering off short-lived 
ground-state constituents, 
subject to euclidean thermodynamics after spatial 
coarse-graining thus is described by a particular radiative correction in the effective 
theory which can be computed in the real-time formulation of 
thermalized quantum field theory and interpreted in purely energetic grounds \cite{LHGS2008}. 

The purpose of the present paper is to go beyond the approximation
$p^2=0$ by numerically computing the function $G$ selfconsistently on 
the radiatively induced mass shell. Qualitatively, all the 
above described, essential features are reproduced by the full calculation:
Im\,$G=0$, screening at low and antiscreening at large spatial 
momentum modulus. At $T\sim 2\,T_c$ there is practically no difference
between the full result and its approximation, and thus 
all results in applications, which rely on the low-temperature behavior 
of $G$, remain valid \cite{SHG2006-2,SH2007,SHGS2007}. The high-temperature
behavior of $G$ is profoundly different in the full calculation: While
the onset of the gap $Y^*=\frac{\omega^*}{T}$ towards small frequencies 
in black-body spectra
falls off at high temperatures as $\lambda^{-2/3}$ in the 
approximate result the full calculation yields a high-temperature
behavior as $Y^*\sim \lambda^{-3/2}$ where $\lambda\equiv\frac{2\pi
  T}{\Lambda}$ is temperature scaled dimensionless by the Yang-Mills
scale $\Lambda$. Thus the cutoff frequency $\omega^*$ absolutely increases
as $\lambda^{-1/2}$ with increasing temperature and not only relatively
to the maximum of the black-body spectrum as was suggested by the
approximate result. 

This article is organized as follows. In Sec.\,\ref{Pi} we first discuss the 
essential steps leading to the effective theory for the deconfining phase of SU(2) 
Yang-Mills thermodynamics and subsequently  remind the
reader of how the part of $\Pi$ associated with the screening function 
$G$ for propagating massless modes is computed. We detail some 
technical aspects in computing $G$ selfconsistently, 
provide our results, and compare them to those of the 
approximate calculation (assuming $p^2=0$ for the 
external four-momentum $p$). This includes a discussion of the 
high-temperature behavior of $G$. We also
stress 
that the function $G$ now saturates to finite values as 
$|\vec{p}|\to 0$. In Sec.\,\ref{BB} we employ the results of 
Sec.\,\ref{Pi} to derive a modification of black-body spectra 
once it is assumed that photon propagation is subject to an SU(2) 
gauge principle. In particular, we point out that in the 
high-temperature regime the approximation 
$p^2=0$ yields differing results compared to the full 
calculation: The Planck spectrum is approached much faster now. 
A summary of our work is given in
Sec.\,\ref{SC}.

\section{Calculation of the 
screening function $G$ on the radiatively induced mass shell \label{Pi}}

\subsection{Essential steps in deriving the effective theory}

Before we address the actual topic of the present work 
we would like to briefly discuss essential steps taken in 
\cite{Hofmann2005,HH2004,Hofmann2007} to arrive at the effective theory for the 
deconfining phase. 

The first step is to derive a thermal ground state composed of topologically 
nontrivial field configurations. The key observation is that BPS saturated, (anti)selfdual 
field configurations in the Euclidean formulation 
exhibit vanishing energy-momentum and thus do not propagate in real time. 
Thus, if a selfconsistent spatial coarse-graining over these configurations can be 
executed then the resulting ground-state describing, adjoint scalar field 
$\phi$ cannot propagate and fluctuate either. In a thermodynamical situation, 
the spatial and temporal homogeneity of one-point 
functions then implies the spatial and temporal homogeneity of $\phi$'s 
modulus $|\phi|$ (inertness). By selfconsistent it is meant that the spatial resolution scale of 
the effective theory emerges in the process of performing the coarse-graining 
in terms of temperature $T$ and an integration constant $\Lambda$. Because of 
$\phi$'s inertness its dynamics is carried by its dimensionless phase $\hat{\phi}$. 
It turns out that on dimensional grounds only one single concrete (nonlocal) definition is possible in terms of BPS saturated 
field configurations describing the kernel of a differential operator which contains 
$\hat{\phi}$. Moreover, because of their sufficiently low-dimensional 
moduli spaces only those configurations possessing topological charge modulus 
$|Q|=1$ may contribute. Finally, only configurations of trivial holonomy may 
contribute since a non-dynamical holonomy parameter, which is required for BPS saturation, 
gives rise to total suppression in the infinite-volume limit \cite{GPY}. This leaves one with 
Harrington-Shepard calorons and anticalorons. 

Executing the above-mentioned, nonlocal definition, which is an average over the 
two-point function of the field strength, points out 
that only magnetic-magnetic correlations do actually invoke a nontrivial result. Moreover, all 
ambiguities arising in the course of the calculations are one-to-one related 
to the undetermined parameters spanning the kernel of the operator 
$D=\partial_\tau^2+\left(\frac{2\pi}{\beta}\right)^2$, where $\beta\equiv 1/T$, which is 
linear. Thus $D\phi=0$ is the sought after equation of motion. 
BPS saturation and this second-order equation are consistent if and only if $\phi$'s 
potential $V(|\phi|^2)$ satifies $\partial_{|\phi|^2} V(|\phi|^2)=-V(|\phi|^2)/|\phi|^2$ \cite{GH2006} with 
solution $V(|\phi|^2)=\Lambda^6/|\phi|^2$ ($\Lambda$ an integration constant of dimension mass). 
Notice that in this way the Yang-Mills scale $\Lambda$ sneaks into the 
game purely nonperturbatively. Now, disregarding any interactions between the topologically 
trivial and the nontrivial sector of the fundamental theory, perturbative renormalizability 
\cite{tHooftVeltman,Zinn-Justin} states that a coarse-graining over 
the former sector doesn't change the {\sl form} of its effective action as compared to 
the fundamental action. The gauge invariance plus the invariance under the allowed 
spacetime symmetries does then only allow for a minimal coupling between $\phi$ and 
the coarse-grained topologically trivial gauge field $a_\mu$, see also \cite{LHGS2008}, 
leading to the following effective action  
\eqb
\label{effdec}
S_{\tiny\mbox{dec}}=\mbox{tr}\,\int d^4x\,\left\{\frac12\,
  G_{\mu\nu}G_{\mu\nu}+D_\mu\phi D_\mu\phi+\frac{\Lambda^6}{\phi^2}\right\}\,.
\eqe
Here $D_\mu$ is an adjoint covariant derivative involving the yet unknown effective 
gauge coupling $e$ and $G_{\mu\nu}=\partial_\mu a_\nu-\partial_\nu a_\mu+ie[a_\mu,a_\nu]$. 
On tree-level, the action in Eq.\,(\ref{effdec}) describes a situation where the thermal 
ground state only knows about the generation of monopoles and antimonmopoles that collapse onto 
each other before they get re-separated through the absorption by (small-holonomy) (anti)calorons 
of propagating fundamental gauge modes that are, compared to the scale $|\phi|$, 
far off their mass shell. The processes associated with a large (temporary) holonomy, leading to (anti)caloron dissociation 
into isolated and screened magnetic monopoles and antimonopoles \cite{Diakonov2004}, collectively are described by 
radiative corrections in the effective theory. Mesoscopically, the field $\phi$ gets domainized by these stable magnetic charges 
\cite{Kibble} (departure from the exponentially fast saturated 
infinite-volume limit considered in deriving $\phi$) but the thermal average 
over these domainizations, which is described by the loop expansion in the effective theory, assures the spatial 
homogeneity of one-point functions. Notice that this loop expansion, which converges rapidly due 
to kinematical constraints imposed by the nontrivial thermal ground state in physical unitary-Coulomb gauge, 
does not count powers of the effective coupling 
$e$. The latter's running with temperature follows from the invariance of Legendre transformations under the applied 
spatial coarse-graining. For $T\gg T_c$ one has $e\equiv\sqrt{8}\pi$ where $T_c$ is the critical temperature for 
the deconfining-preconfining transition (more precisely: coexistence of these phases).

\subsection{Set-up and computation of real part of $\Pi$}

\begin{figure}
\begin{center}
\leavevmode
\leavevmode
\vspace{4.9cm}
\includegraphics{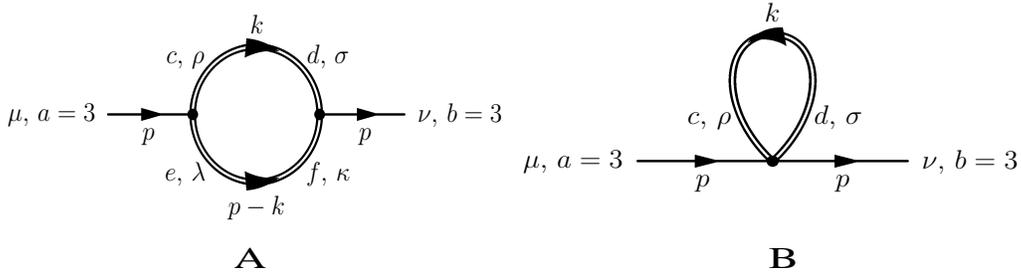}
\end{center}
\caption{\protect{\label{Fig-1}} 
The Feynman diagrams for the massless mode's polarization tensor. 
Double (single) lines denote propagators for the tree-level 
massive (tree-level massless) mode. }      
\end{figure}

In Sec.\,3 of  \cite{SHG2006} we have given the Feynman rules in
unitary-Coulomb gauge for the
computation of radiative corrections in the effective theory for the
deconfining phase of SU(2) Yang-Mills thermodynamics 
\cite{Hofmann2005}. For the sake brevity we 
do not repeat them here. On the one-loop level
the two diagrams contributing to the polarization tensor
$\Pi^{\mu\nu}(p,T)$ of the massless 
mode of four-momentum $p$ are shown in Fig.\,\ref{Fig-1}. 
Thus $\Pi^{\mu\nu}=\Pi_A^{\mu\nu}+\Pi_B^{\mu\nu}$ where  
\begin{equation}
\label{Anproc}
\begin{split}
\Pi^{\mu\nu}_{A}(p)\,=\,& 
\frac{1}{2i}\int\frac{d^4k}{(2\pi)^4} e^2 
	\epsilon_{ace}[g^{\mu\rho}(-p-k)^\lambda+g^{\rho\lambda}(k-p+k)^\mu+g^{\lambda\mu}(p-k+p)^\rho]\times\\
& \epsilon_{dbf}[g^{\sigma\nu}(-k-p)^\kappa+g^{\nu\kappa}(p+p-k)^\sigma+g^{\kappa\sigma}(-p+k+k)^\nu]\times\\
&(-\delta_{cd})\left(g_{\rho\sigma}-\frac{k_\rho k_\sigma}{m^2}\right) 
\left[\frac{i}{k^2-m^2}+2\pi\delta(k^2-m^2)\,n_B(|k_0|/T) \right]\times\\
&(-\delta_{ef})\left(g_{\lambda\kappa}-\frac{(p-k)_\lambda(p-k)_\kappa}{(p-k)^2}\right)\times\\
&\left[\frac{i}{(p-k)^2-m^2}+2\pi\delta((p-k)^2-m^2)\,n_B(|p_0-k_0|/T) \right]
\end{split}
\end{equation}
and 
\begin{equation}
\label{vactad}
\begin{split}
\Pi^{\mu\nu}_{B}(p)\,=\,&\frac{1}{i} \int \frac{d^4k}{(2\pi)^4} 
(-\delta_{ab}) \left( g_{\rho\sigma}-\frac{k_\rho k_\sigma}{m^2} \right)  
\left[\frac{i}{k^2-m^2}+2\pi\delta(k^2-m^2)n_B(|k_0|/T) \right]\times\\
 & \quad (-ie^2)[
 \epsilon_{abe}\epsilon_{cde}(g^{\mu\rho}g^{\nu\sigma}-g^{\mu\sigma}g^{\nu\rho})
 +\epsilon_{ace}\epsilon_{bde}(g^{\mu\nu}g^{\rho\sigma}-g^{\mu\sigma}g^{\nu\rho})+\\
 &\quad\epsilon_{ade}\epsilon_{bce}(g^{\mu\nu}g^{\rho\sigma}-g^{\mu\rho}g^{\nu\sigma})]\,,
\end{split}
\end{equation}
where $m\equiv 2e|\phi|$ is the common mass of the two vector modes induced by the
adjoint Higgs mechanism in the effective theory and
$|\phi|\equiv \sqrt{\frac{\Lambda^3}{2\pi T}}$. The relation between 
Yang-Mills scale $\Lambda$ and critical temperature $T_c$ for the
deconfining-preconfining phase transition is
$\Lambda=\frac{2\pi}{13.87}\,T_c$ \cite{Hofmann2005}. 
It was shown in \cite{SHG2006} that $\Pi_A^{\mu\nu}\equiv 0$ for 
$p^2=0$. We will see below that this continues to hold 
on the radiatively induced mass shell so that 
$\Pi^{\mu\nu}$ is always real.  

Without constraining generality we assume $\vec{p}$ to be parallel to
the $z$-axis. In this case, the screening function 
$G(p_0(\vec{p}),\vec{p}))$, which enters the radiatively 
induced dispersion law as 
\eqb
\label{displaw}
p_0^2(\vec{p})=\vec{p}^2+G(p_0(\vec{p}),\vec{p})\,,
\eqe
is given as 
\eqb
\label{gxxgyy}
\Pi^B_{11}=\Pi^B_{22}=G(p_0,\vec{p})\,.
\eqe
The integration in Eq.\,(\ref{vactad}) is subject to constraints
imposed by the existence of the thermal ground state in the effective
theory. Namely, the momentum transfer in the four-vertex of diagram B in 
Fig.\,\ref{Fig-1} satisfies the following condition
\begin{equation}
\label{condproc}
\begin{split}
|(p+k)^2|=\left|G+2\,k\cdot p+ 4e^2|\phi|^2\right|\leq |\phi|^2\,,
\end{split}
\end{equation}
where $|\phi|=\sqrt{\frac{\Lambda^3}{2\pi T}}$. 
Notice that in a selfconsistent calculation of $G$ on the radiatively
induced mass shell in performing the square 
on the left-hand side of inequality 
(\ref{condproc}) the following relations hold
\eqb
\label{squares}
p^2=G\,,\ \ k^2=4e\,|\phi|^2\,,
\eqe
where $e$ is the effective gauge coupling 
whose high-temperature value is given as $e=\sqrt{8}\pi$
\cite{Hofmann2007,GH2007}. 

\begin{figure}
\begin{center}
\leavevmode
\leavevmode
\vspace{5.8cm}
\includegraphics{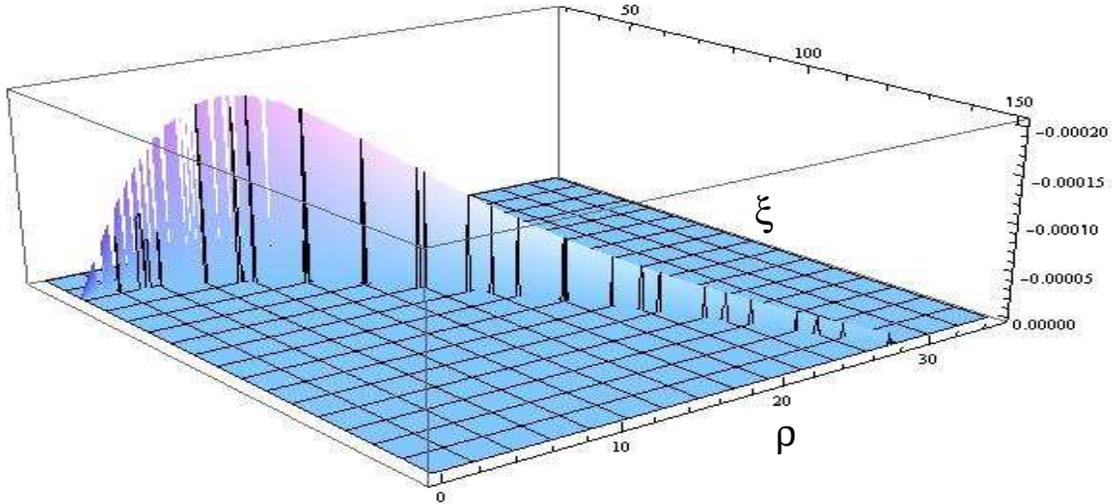}
\end{center}
\caption{\protect{\label{Fig-2}} The integrand in Eq.\,(\ref{intlimts}) subject to the 
constraints in Eq.\,(\ref{constcyl}) for $\lambda=50$, $X=0.8$, $\frac{G}{T^2}=-0.00146$, and
$0\le\rho\le 35$, $30\le\xi\le 150$. 
This example illustrates the considerable constraining power of 
(\ref{constcyl}), here in the regime of antiscreening.}
\end{figure}

In going over to dimensionless variables 
\eqb
\label{vardimless}
\vec{y}\equiv\frac{\vec{k}}{|\phi|}\, 
\eqe
and by transforming to cylindrical coordinates \cite{SHG2006} 
\eqb
\label{cylcoo}
y_1=\rho\,\cos\varphi\,,\ \ \ \ y_2=\rho\,\sin\varphi\,,\ \ \ \ y_3=\xi\,.
\eqe
we have\footnote{The contribution of the vacuum parts in the 
propagators of the two tree-level massive modes is excluded 
by the constraints on the maximal resolution in the 
effective theory, see \cite{Hofmann2005}.}  
\eqb
\label{intlimts}
\frac{G}{T^2}=\int d\xi\,\int d\rho\, 
e^2\lambda^{-3}\left(-4+\frac{\rho^2}{4e^2}\right)\,\rho\,
\frac{n_B\left(2\pi \lambda^{-3/2}\sqrt{\rho^2+\xi^2+4e^2}\right)}
{\sqrt{\rho^2+\xi^2+4e^2}}\,,
\eqe
where $n_B(x)=1/(e^x-1)$. In addition, we define $X\equiv\frac{|\vec{p}|}{T}$. 
The limits of integration in
Eq.\,(\ref{intlimts}) need yet to be determined in accord 
with Eq.\,(\ref{condproc}). In cylindrical coordinates $\rho,\xi$ the support of the integration in Eq.\,(\ref{intlimts}) is the region where $\rho$ and 
$\xi$ satisfy one of the two following conditions 
\eqb
\label{constcyl}
\left|\frac{G}{T^2}\frac{\lambda^{3}}{(2\pi)^2}\pm\frac{\lambda^{3/2}}{\pi}\left(\sqrt{X^2+
\frac{G}{T^2}}\sqrt{\rho^2+\xi^2+4e^2}-X\xi\right)+4e^2\right|\le 1\,.
\eqe
In Fig.\,\ref{Fig-2} the integrand in Eq.\,(\ref{intlimts}) subject to the 
constraints in Eq.\,(\ref{constcyl}) is plotted for
$\lambda=2\,\lambda_c$ and $\lambda=4\,\lambda_c$ where
$\lambda_c=13.87$ is the dimensionless critical temperature for the
deconfining-preconfining phase transition signalled by a logarithmic
pole in $e(\lambda)$ \cite{Hofmann2005,Hofmann2007}.
Through the conditions (\ref{constcyl}) the right-hand side of Eq.\,(\ref{intlimts}) becomes a 
function of $G$ in contrast to the approximate calculation where $p^2=0$. The strategy to determine $G$ at a given value of $\lambda$ 
in the full calculation 
thus is to prescribe a value for $G$ in (\ref{constcyl}), subsequently to compute the integral by Monte-Carlo methods, and to 
list this integral as a function of $G$. The final step is to determine the zero of left-hand side minus 
right-hand side of Eq.\,(\ref{intlimts}) to find $G$ selfconsistently. Numerically, we use Newton's 
method for this task.  
  
\begin{figure}
\begin{center}
\leavevmode
\leavevmode
\vspace{5.8cm}
\includegraphics{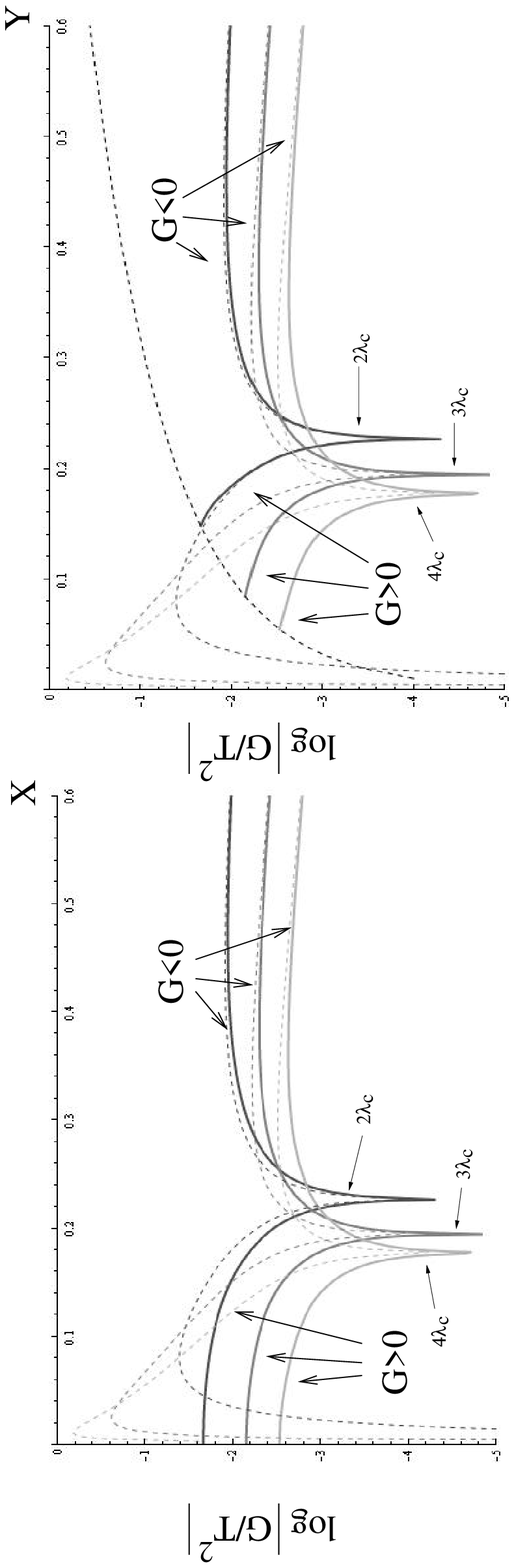}
\end{center}
\caption{\protect{\label{Fig-3}} Plots of $\log\left|\frac{G}{T^2}\right|$ in the full calculation 
(solid grey curves) and for the approximation $p^2=0$ (dashed grey
curves). The cusps in $\log\left|\frac{G}{T^2}\right|$ correspond to
zeros separating the regime of screening ($G>0$) from the regime of
antiscrenning ($G<0$). The left panel depicts 
$\log\left|\frac{G}{T^2}\right|$ as a function of $X$. 
The right panel shows $\log\left|\frac{G}{T^2}\right|$ as a function of $Y\equiv\sqrt{X^2+\frac{G}{T^2}}$. 
Here the dashed black curve is the function $2\log Y$. In order of 
increasing lightness the curves correspond to $\lambda=2\,\lambda_c$, $\lambda=3\,\lambda_c$, and $\lambda=4\,\lambda_c$.}
\end{figure}

In Fig.\,\ref{Fig-3} we compare plots of $\log\left|\frac{G}{T^2}\right|$ obtained in
the full calculation with those of the approximation $p^2=0$, both as a 
function of $X$ and $Y\equiv\sqrt{X^2+\frac{G}{T^2}}$ at 
$\lambda=2\,\lambda_c$, $\lambda=3\,\lambda_c$, and $\lambda=4\,\lambda_c$. Notice the saturation of 
$\frac{G}{T^2}$ in the full calculation to finite values as $X\to 0$. This is in contrast to the 
result obtained in the approximation $p^2=0$ and an important observation since it is likely to  
explain the quadratic rise of the spatial string tension with high temperature in terms of a 
radiative effect in the effective theory for deconfining SU(2)
Yang-Mills thermodynamics \cite{LHGS2008}. Notice also 
the more rapid approach to zero of 
the critical value $Y^*$ for total screening (intersection point of $2\log Y$ and $\log\frac{G}{T^2}(Y)$) in the full 
calculation\footnote{Modes of frequency less than $Y^*$ do no longer propagate.}. 
The regimes of screening (left of the zero of $\frac{G}{T^2}$) and antiscreening  
(right of the zero of $\frac{G}{T^2}$) are the same for the 
full and approximate calculation.   
Finally, let us investigate the high-temperature behavior of $Y^*(\lambda)=\frac{G}{T^2}(X=0,\lambda)$ 
which cuts off the spectrum towards low frequencies.  
\begin{figure}
\begin{center}
\leavevmode
\leavevmode
\vspace{6.8cm}
\includegraphics{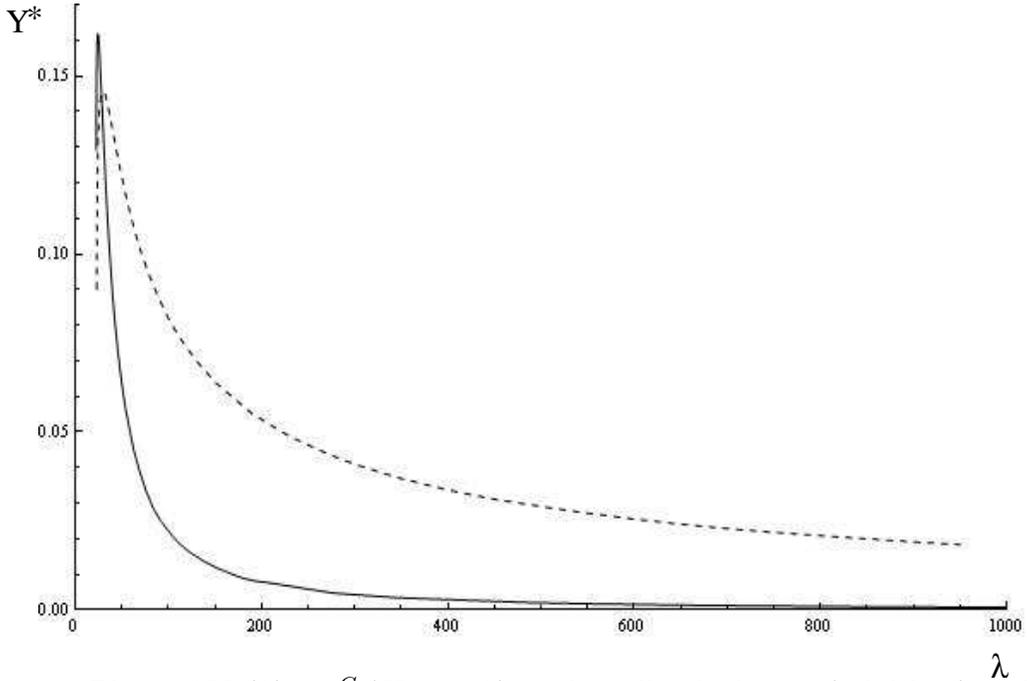}
\end{center}
\caption{\protect{\label{Fig-4}} Plots of 
$Y^*(\lambda)=\frac{G}{T^2}(X=0,\lambda)$ in the full calculation (solid line) and in 
the approximation $p^2=0$ (dashed line).}
\end{figure}
We fit the high-temperature behavior of $Y^*(\lambda)$ to a power-law model
\eqb
\label{hTpl}
Y^*(\lambda)=C\lambda^\nu\,,\ \ \ \ \ \ (\lambda\gg\lambda_c)\,,
\eqe
where $C$ and $\nu$ are real constants. For the full (approximate)
result we obtain 
$C\sim 20$ and $\nu\sim -3/2$ 
($C\sim 2$ and $\nu\sim -2/3$), see also Fig.\,\ref{Fig-4}. 
Thus the approach to abelian behavior is much faster in the 
full calculation as compared to the approximate result.  

\subsection{Check that Im\,$G=0$}

So far we have considered diagram B in Fig.\,\ref{Fig-1}. Keeping
in mind that only the thermal parts of the massive vector 
propagators contribute in diagram A \cite{Hofmann2005} and by 
inspecting the analytic expression in Eq.\,(\ref{Anproc}), it is clear
that if this diagram is finite then it necessarily is 
purely imaginary. Taking into account that $p^2=G$, we define 
in dimensionless cartesian coordinates the following 
functions
\eab
\label{defsA}
y_3(j)&\equiv&-\frac{\lambda^{3/2}T^2}{4\pi
  G}\left(-X\frac{G}{T^2}+\right.\nonumber\\ 
&&\left.(-1)^j\sqrt{\left(\frac{XG}{T^2}\right)^2
+4\,\frac{G}{T^2}\left[\frac{G^2}{4T^4}-2\pi\lambda^{-3/2}(X^2+\frac{G}{T^2})(y_1^2+y_2^2+4e^2)\right]}\right)\,,
\nonumber\\ 
s(j)&\equiv&\sqrt{y_1^2+y_2^2+y_3(j)^2+4e^2}\,,\nonumber\\ 
r(i,j)&\equiv&2\pi\lambda^{-3/2}\left((-1)^i\,s(j)\sqrt{X^2+\frac{G}{T^2}}-
  2\pi\lambda^{-3/2}y_3(j)X\right)\,,\nonumber\\ 
\eae
where $i,j=0,1$. Then we have 
\eab
\label{expreA}
\frac{\Pi^A_{11}}{T^2}&=&\frac{\Pi^A_{22}}{T^2}=\frac{ie^2}{(2\pi)^2}\sum_{i,j=0}^1
\int_{-\infty}^{\infty}dy_1\int_{-\infty}^{\infty} dy_2\,,\nonumber\\ 
&&\left\{\left[-2\,r(i,j)-\left(\frac{\lambda^{3/2}}{2\pi}\right)^2\,\frac{r(i,j)^2}{4e^2}+
7\,\frac{G}{T^2}-\left(\frac{\lambda^{3/2}}{2\pi}\right)^2\left(\frac{G}{2eT^2}\right)^2\right]+
\right.\nonumber\\ 
&&\left.\left[48\,\pi^2\lambda^{-3}-2\,\frac{r(i,j)}{4e^2}-
3\,\frac{G}{4e^2\,T^2}+\left(\frac{\lambda^{3/2}}{2\pi}\right)^2
\left(\frac{G}{4e^2\,T^2}\right)^2\right] y_1\,y_2
\right\}\times\nonumber\\  
&&\frac{\pi}{\left|2\,\lambda^{3/2}y_3(j)\sqrt{X^2+\frac{G}{T^2}}\right|}
n_B(2\pi\lambda^{-3/2}s(j))\,n_B\left(\left|\sqrt{X^2+\frac{G}{T^2}}+(-1)^i2\pi
\lambda^{-3/2}s(j)\right|\right)\,,\nonumber\\ 
\eae
and no further constraints need to be imposed. We have computed the integral 
in Eq.\,(\ref{expreA}) employing Monte-Carlo methods and by using our result for 
$G$ (determined by $\Pi_{11}^B$) for various temperatures 
$\lambda$. We constantly obtain a vanishing result. Thus there is no finite photon 
width, and the problem of computing $G$ is selfconsistently 
solved by computing $\Pi_{11}^B$ only.

\section{Modified black-body spectra\label{BB}} 

Here we would like to compare the implication of the full one-loop 
result for $G$ for black-body spectra with that of the approximation $p^2=0$ under the postulate 
that photon propagation is described by the propagation of the massless mode in the deconfining phase 
of an SU(2) Yang-Mills theory  (SU(2)$_{\tiny\mbox{CMB}}$) 
of scale $\Lambda\sim 10^{-4}\,$eV which corresponds to $T_c=2.73\,$K 
\cite{Hofmann2005}.  

\begin{figure}
\begin{center}
\leavevmode
\leavevmode
\vspace{5.8cm}
\includegraphics{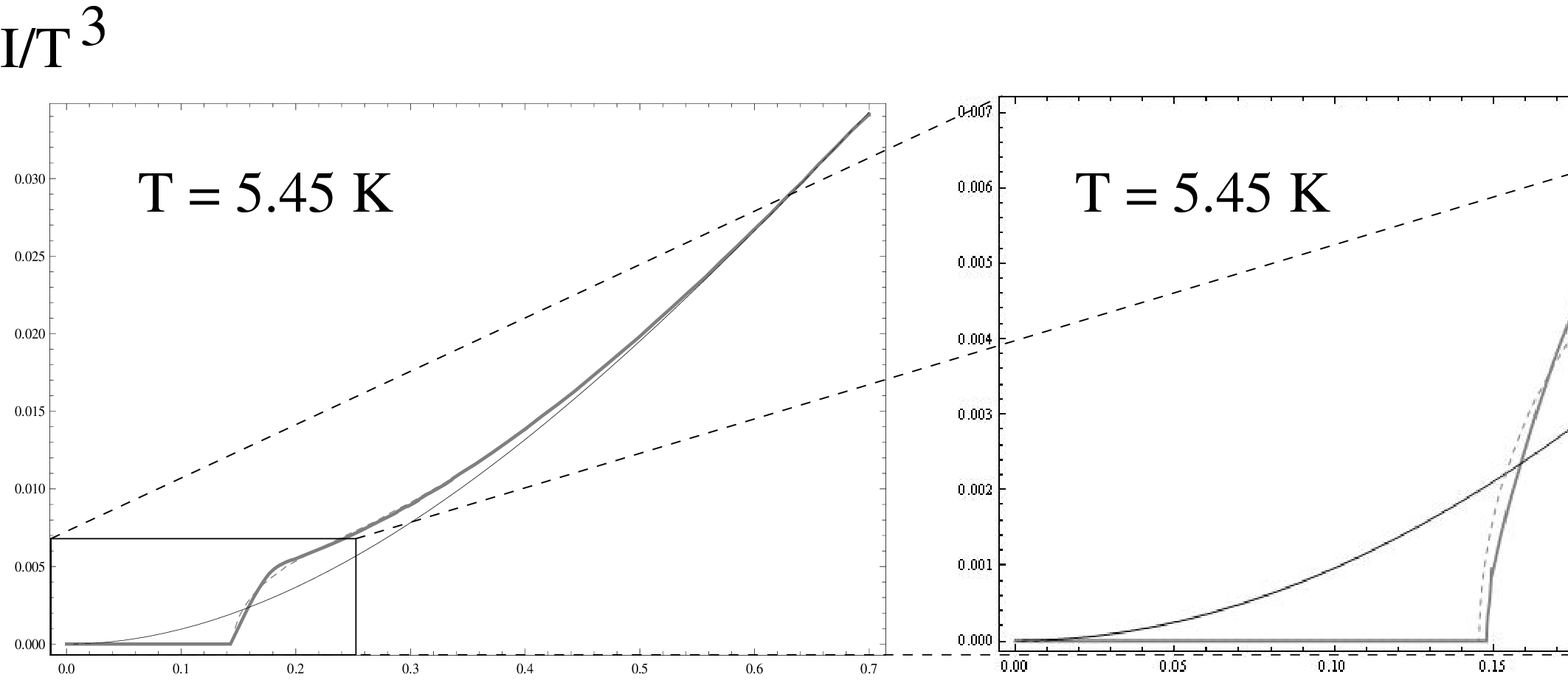}
\end{center}
\caption{\protect{\label{Fig-5}} Plots of the dimensionless spectral 
intensity $\frac{I}{T^3}$ as a result of the full calculation 
(solid grey curves) and for the approximation $p^2$ (dashed grey
curves). 
Solid black curves depict the conventional Planck 
spectrum for $T=2\,T_c\sim 5.45\,K$. Notice the excellent agreement with
the result of the approximate calculation.}
\end{figure}

\begin{figure}
\begin{center}
\leavevmode
\leavevmode
\vspace{5.8cm}
\includegraphics{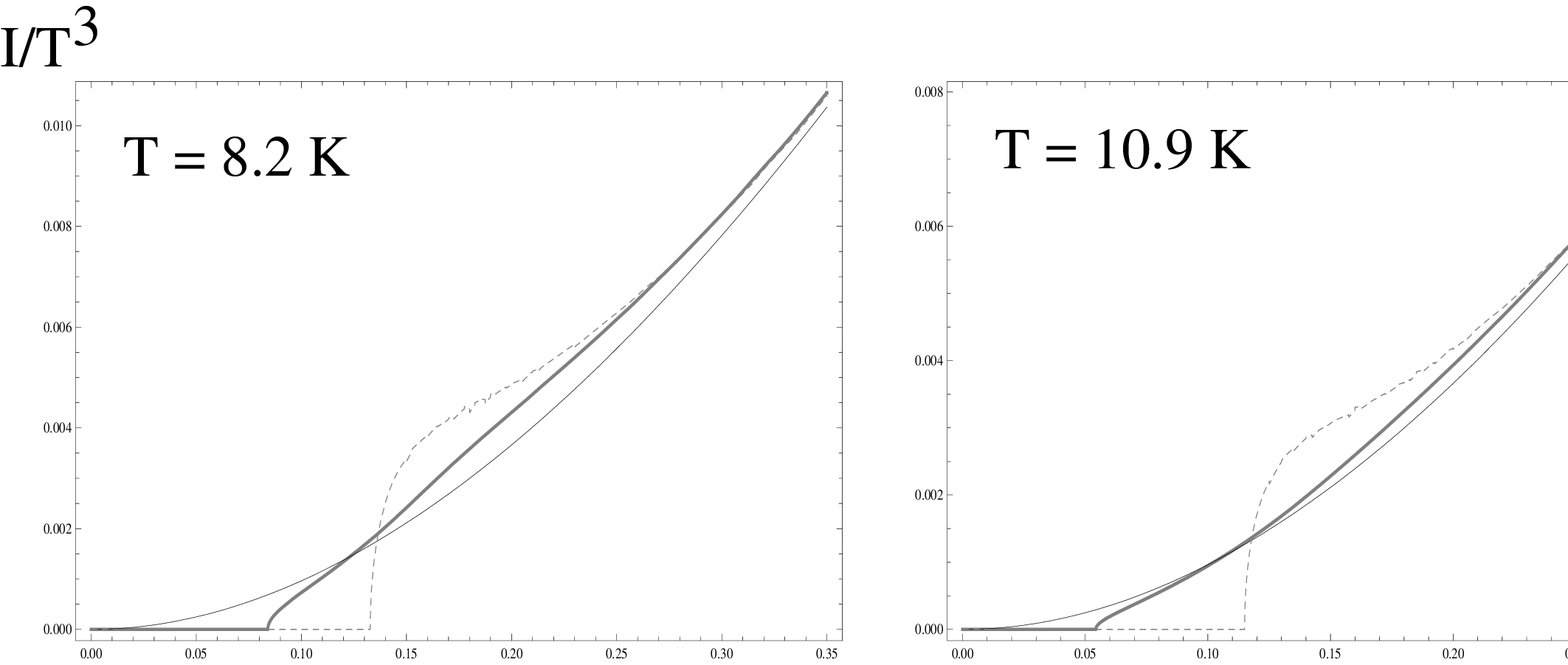}
\end{center}
\caption{\protect{\label{Fig-6}} Plots of the dimensionless spectral 
intensity $\frac{I}{T^3}$ in the full calculation 
(solid grey curves) and for the approximation $p^2$ (dashed grey
curves). 
Solid black curves depict the conventional Planck 
spectrum for (a) $T=3\,T_c\sim 8.2\,K$ and (b) $T=4\,T_c\sim 10.9\,K$. 
Notice the much faster approach with rising temperature to the Planck 
spectrum in the full as compared to the approximate calculation.} 
\end{figure}

As was explained in \cite{SHG2006-2,RH2007} the modified black-body spectral intensity $I_{\tiny\mbox{SU(2)}}(\omega)$ 
of the SU(2) theory is obtained from the conventional counterpart $I_{\tiny\mbox{U(1)}}(\omega)$ of the
 U(1) theory as follows  
\eqb
\label{PmBB}
I_{\tiny\mbox{U(1)}}(\omega)\to I_{\tiny\mbox{SU(2)}}(\omega)=I_{\tiny\mbox{U(1)}}(\omega)\times
\frac{\left(\omega-\frac{1}{2}\frac{d}{d\omega}G\right)\sqrt{\omega^2-G}}{\omega^2}\,
\theta(\omega-\omega^*)\,
\eqe
where $\omega^*\equiv\sqrt{G(\vec{p}=0,T)}$,  
\eqb
\label{PiBB}
I_{\tiny\mbox{U(1)}}(\omega)\equiv\frac{1}{\pi^2}\,\frac{\omega^3}{\exp[\frac{\omega}{T}]-1}\,,
\eqe
and $\theta(x)$ is the Heaviside step 
function. In Fig.\,\ref{Fig-5} a comparison of the modification of low-frequency black-body intensities 
obtained in the full calculation and in the approximation $p^2=0$ is 
depicted for the temperature $T=2\,T_c\sim 5.45$\,K. Notice the
excellent agreement of the approximate and the exact result. 
In Fig.\,\ref{Fig-6} this comparison is made for temperatures 
$T=3\,T_c\sim 8.2$\,K and $T=4\,T_c\sim 10.9$\,K. 
Notice the faster approach to the Planck spectrum with rising
temperature in the full as compared to the approximate
calculation. 

\section{Summary and Conclusions\label{SC}}

Let us summarize our results. In \cite{SHG2006} the 
components of the polarization tensor of the massless mode (photon) were
computed. These are relevant 
for the radiatively induced dispersion law in the effective theory for
the deconfining phase of SU(2) Yang-Mills thermodynamics. 
To make the calculation as analytically transparent as possible 
(resolution of kinematic constraints as imposed by the nontrivial thermal
ground state) and also out of convenience, reference \cite{SHG2006} has 
resorted to the approximation that the photon's 
four momentum be on its free mass shell ($p^2=0$). Under the assumption that
physical photon propagation is described by an SU(2) 
Yang-Mills theory of scale $\Lambda\sim 10^{-4}\,$eV in 
\cite{SHG2006-2} implications were investigated for the low-frequency part of
low-temperature black-body spectra obtaining a sizable gap and a regime
of sizable antiscreening. 

The calculation of the photon's polarization tensor as performed 
in the present paper no longer resorts to the above approximation. 
Rather, the incoming photon is now placed selfconsistently on its
radiatively induced mass shell. For black-body spectra we obtain excellent 
agreement with the approximate calculation for temperatures up to
$\sim$\,6\,Kelvin. 
At higher temperatures the one-loop exact shows a less
dramatic effect than the approximate calculations suggests. Namely, 
the black-body spectrum in the low-frequency part of the antiscreening
regime is less bumpy, and the
spectral gap decreases faster with temperature. 

Finally, notice that since results obtained in the approximation 
$p^2=0$ yield agreement with the full calculation up to $T\sim 6\,$ 
Kelvin and since in applications \cite{SH2007,SHGS2007} we see an effect 
only for temperatures up to this size the conclusions spelled out in these papers 
remain quantitatively valid. 

\section*{Acknowledgments}
The authors gratefully acknowledge fruitful discussions with Francesco Giacosa,
Jochen Keller, and Markus Schwarz. We also would like to thank Markus
Schwarz for his helpful comments on the manuscript.

\baselineskip25pt

\end{document}